\documentclass[12pt,amsfonts]{article}

 \usepackage{amssymb}
 \def\N{{\cal{N}}}
\def\gg{\hat{g}}

\def\half{\textstyle{\frac{1}{2}}}

\def\H{{\cal H}}

\def\th{\theta}
\def\H{{\cal H}}

\def\D{{\cal D}}

\def\ra{\rightarrow}
\def\tint{{\textstyle\int}}
\def\hg{{\hat g}}
\def\hp{{\hat\pi}}

\def\s{\hskip.08em}
\def\d{\partial}
\def\Op{{\cal{O}}}
\def\ol{\overline}

\def\b{\begin{eqnarray*}}  
\def\e{\end{eqnarray*}}    
\def\bn{\begin{eqnarray}}  
\def\en{\end{eqnarray}}   

\def\<{\langle}
\def\>{\rangle}

\def\no{\nonumber}

\def\{{\lbrace}

\def\}{\rbrace}
\begin{document}

\title{ Let Loop Quantum Gravity \\ and Affine Quantum Gravity\\ Examine Each Other}
  \author{John R. Klauder\footnote{klauder@ufl.edu} \\
Department of Physics and Department of Mathematics  \\ 
University of Florida,   
Gainesville, FL 32611-8440}
\date{ }
\bibliographystyle{unsrt}

\maketitle 

\begin{abstract} Loop Quantum Gravity is widely developed using canonical quantization in an effort to find the correct
quantization for gravity. Affine quantization, which is like canonical quantization augmented 
bounded in one orientation, e.g., a strictly positive coordinate. 
We open discussion using canonical and affine quantizations for two simple problems so each procedure can be understood. That analysis opens a modest treatment of quantum gravity gleaned from some   
typical features that exhibit the profound differences between aspects of seeking the quantum  treatment of Einstein's gravity.
\end{abstract}

\section{Introduction} 
We begin with two different quantization procedures, and two simple, but distinct, problems one of which is successful and the other one is a failure using both of the quantization procedures. 

This exercise serves as a prelude to a valid and straightforward quantization of gravity

\subsection{Choosing a canonical quantization}
The classical variables, $p\&q$, that are elements of a constant zero curvature, better known as Cartesian variables, such as those featured by Dirac \cite{dirac}, are promoted to self-adjoint quantum operators $P\,(=P^\dag)$ and
 $Q\,(=Q^\dag)$, ranged as $-\infty<P,Q<\infty$, and scaled so that $[Q, P]=i\hbar \,1\!\!1$.{\footnote{In particular, in \cite{dirac}, the mid-page of 114, Dirac wrote ``However, if the system does have a classical analogue, its connexion with classical mechanics is specially close and one can usually assume that
the Hamiltonian is the same function of the canonical coordinates and momenta in the quantum theory
as in the classical theory.$\dag$ "
Footnote ``$\dag$  This assumption is found in practice to be successful only when applied
with the dynamical coordinates and momenta referring to a Cartesian system of axes and not
to more general curvilinear coordinates."} 

\subsubsection{First canonical example}
Our example is just the familiar harmonic oscillator, for which $-\infty<p,q<\infty$ and a Poisson bracket $\{q,p\}=1$, chosen with a parameter-free, classical Hamiltonian, given by $H(p,q) = (p^2+q^2)/2$. The quantum Hamiltonian is ${\cal{H}}(P,Q)=(P^2+Q^2)/2$, and Schr\"odinger's representation is given by $P=-i\hbar(\d/\d x)$ and $Q=x$. Finally, for our example, Schr\"odinger's equation is given by
    \bn i\hbar(\d \psi(x,t)/\d t)= ( -\hbar^2 (\d^2/\d x^2)+ x^2)/2\;\psi(x,t)\;. \label{1} \en
    
    Solutions to Eq.~(\ref{1}) for our example are well known. In particular, for the harmonic oscillator: the eigenvalues are given by 
    $E_n = \hbar(n + 1/2)$ for $n=0,1,2,,...$; the eigenfunctions (with $\hbar=1$) are given by $\psi_n(x)=N\,H_n(x) \,e^{-x^2/2}$ with $ n=0,1,2,...$; $H_n(x) $ are the  Hermite polynomials  \cite{2}; and $N$ is a normalization factor.
    
    \subsubsection{Second canonical example}
    For our next example
     we keep the same classical Hamiltonian, and we retain  $-\infty<p<\infty$, but now we restrict $0<q<\infty$. This new model is called a half-harmonic oscillator. It follows that the operator 
    $P\neq P^\dag$, which leads to a  different behavior than when $P$ was self adjoint. In particular, we now have two distinct quantum Hamiltonians, specifically ${\cal{H}}_0(P,Q)=(PP^\dag + Q^2)/2$, while the other is ${\cal{H}}_1(P,Q)=(P^\dag P+Q^2)/2$. Each of these quantum Hamiltonians lead to the same classical Hamiltonian, namely $(p^2+q^2)/2$, when $\hbar\ra 0$.
 Hence, there are two valid sets of eigenvalues, specifically  $E_0= \hbar[(0,2,4,...)+1/2]$ and $E_1=\hbar
 [(1,3,5,...)+1/2]$. Moreover, these two eigenvalues can be weaved together, such as  $E = \hbar[(0,2,3,5,7,8,...)+1/2]$, which implies that there are {\it infinitely many versions of the spectrum}.
 
 That judgement renders canonical quantization of the half-harmonic oscillator to be an invalid quantization.
 
 \subsubsection{First affine example}
 The traditional classical affine variables are $d\equiv pq$ and $q>0$, and they have a Poisson bracket given by
 $\{q,d\}=q$. 
 Although the harmonic oscillator extends over the whole real line, we choose $q$ to satisfy 
 $-b<q<\infty$, with $b>0$. For very large $b$ we can approximate a full-line harmonic oscillator
 and even see what happens if we choose $b\ra \infty$ to mimic the full-line story.
 
 The classical affine variables now are $-\infty< d\equiv p(q+b)<\infty$ and $0<(q+b)<\infty$, while the classical harmonic oscillator Hamiltonian is given by $H'(d,q) = [d^2\,(q+b)^{-2}+ q^2]/2$, an
 expression that obeys $H(p,q)=(p^2+q^2)/2$ albeit that $-b<q<\infty$.
 
 Now we consider basic quantum operators, namely  $D=[P^\dag (Q+b)+ (Q+b)P)]/2$ and $Q+b$, which lead to $[Q+b, D]=i\hbar\,(Q+b)$, along with $Q+b>0$. The quantum partial-harmonic oscillator is now given by
   \bn  &&H'(D,Q) =[D(Q+b)^{-2}D+ Q^2]/2\;, \en
   which, when $b\ra\infty$, we find $\lim_{b\ra\infty} H'(D,Q) =(P^\dag P + Q^2)/2)$, for which $ P^\dag=P$ now since  $-\infty<Q<\infty$. Furthermore, $\lim_{b\ra\infty} [Q+b, D++bP]/b =
   \lim_{b\ra\infty} [Q, D+bP]/b =[Q,P]=i\hbar \,1\!\!1$. 
   
   In short, an affine quantization {\it becomes} a canonical quantization when the partial real line 
   ($-b<q,Q<\infty$) is stretched to its full length,
    $(-\infty<q,Q<\infty)$. 
    
    Evidently, an affine quantization fails to quantize a full harmonic oscillator.

 \subsubsection{Second affine example}
 The common canonical operator expression, $[Q,P]=i\hbar,1\!\!1$, {\it directly implies} $[Q,(PQ+QP)/2]=i\hbar\,Q$, {\it the basic affine operator!}. 
 To confirm  the affine expression,
 multiply $i\hbar1\!\!1=[Q,P]$ by $Q$, which gives $i\hbar\,Q=(Q^2P-PQ^2)/2
 =(Q^2P+QPQ - QPQ-PQ^2)/2$, i.e., $i\hbar \,Q=[Q,(QP+PQ)/2]$, which is the basic affine expression, $[Q,D]=i\hbar \,Q$, where $D\equiv (PQ+QP)/2$. {\it This derivation {\bf demands} that $Q>0$ and/or $Q<0$. Canonical quantization implies affine quantization, but adds a limit on coordinates.}\footnote{
  A multiplication by $Q^2-1$, instead of $Q$, could create a {\it finite range of coordinates}.}}

Regarding our problem, now $b=0$, and so the classical affine variables are $d\equiv pq$ and $q>0$, which lead to the    half-harmonic oscillator $H'(d ,q)=(d^2\,q^{-2} + q^2)/2$. The basic affine quantum operators are                                    $D$ and $Q>0$, where $ (P^\dag Q+QP)/2 \equiv D\,(=D^\dag)$ and $Q>0\,(=Q^\dag>0)$. 
 These quantum variables lead to $[Q,D]=i\hbar\,Q$.
 The half-harmonic oscillator quantum Hamiltonian is given by
 $H'(D,Q)=(DQ^{-2}D + Q^2)/2$. Schr\"odinger's representation is given by $Q=x>0$ and
 \bn D=-i\hbar [x(\d/\d x)+(\d/\d x)x) ]/2= -i\hbar[x(\d/\d x)+1/2] \;.\en
 Finally, Schr\"odinger's equation is given by
     \bn &&\hskip-3em i\hbar(\d \psi(x,t)/\d t =\label{33} \\ &&\hskip-2em =
      [-\hbar^2(x(\d/\d x)+1/2)\,x^{-2}\,(x(\d/\d x)+1/2) + x^2]/2\;\psi(x,t) \no \\
     &&\hskip-2em= [-\hbar^2 \,(\d^2/\d x^2)+(3/4)\hbar^2/x^2 +x^2]/2 \;\psi(x,t)\;. \no \en
     
  Solutions of (\ref{33}) have been provided by L. Gouba \cite{lg}. Her solutions for the half-harmonic oscillator contain eigenvalues that are equally spaced as are the eigenvalues of the full-harmonic oscillator, although the spacing itself differs in the two cases. The relevant differential equation in (\ref{33}) is known as a spiked harmonic oscillator, and its solutions are based on confluent hypergeometric functions. It is noteworthy that every eigenfunction $\psi_n(x)\varpropto x^{3/2}$, when $0<x\lll 1$, for all $n=0,1,2,...$.
  
  Finally, an affine quantization of the half-harmonic oscillator can be considered a correctly solved problem.
  
\subsection{Lessons from canonical and affine \\quantization procedures}
 An important lesson from the foregoing set of examples is that canonical quantization requires classical variables, i.e., $p\&\,q$, to be promoted to quantum operators must satisfy $-\infty < p,q<\infty$. However, an affine 
 quantization requires classical variables, i.e., $d\& q$, to be promoted to quantum operators must satisfy $-\infty<d<\infty$ and  $-b< \pm q<\infty$, provided that $b$ is finite.
 
 
 Our analysis of gravity will exploit the fact that the metric (the analog of $q$) is required to be {\it strictly positive}, i.e., $ds(x)^2=g_{ab}(x)\,dx^a\,dx^b>0$, provided $\{dx^a\} \equiv\hskip-1.05em/\;\;0$. In addition, the metric itself, $g_{ab}(x)$, is dimensionless.
 These important properties are well adapted to the possibilities offered by affine quantization.
 
 \section{A Loop Quantum Gravity for \\Einstein's Gravity}
 To begin this section the author has chosen to offer the initial paragraph regarding ``Loop Quantum Gravity''
 from Wikipedia in order to have a common text provided by proponents  of that topic.\footnote{The references within this paragraph are those belonging to Wikipedia and are not references chosen by the author for the rest of this article.}
 
  \subsection{The initial paragraph from Wikipedia's \\``Loop Quantum Gravity''}
 ``Beyond the Standard Model
CMS Higgs-event.jpg
Simulated Large Hadron Collider CMS particle detector data depicting a Higgs boson produced by colliding protons decaying into hadron jets and electrons
Standard Model
show
Evidence
show
Theories
show
Supersymmetry
hide
Quantum gravity
False vacuum String theory Spin foam Quantum foam Quantum geometry Loop quantum gravity Quantum cosmology Loop quantum cosmology Causal dynamical triangulation Causal fermion systems Causal sets Canonical quantum gravity Semiclassical gravity Superfluid vacuum theory
show
Experiments
vte
Loop quantum gravity (LQG)[1][2][3][4][5] is a theory of quantum gravity, which aims to merge quantum mechanics and general relativity, incorporating matter of the Standard Model into the framework established for the pure quantum gravity case. As a candidate for quantum gravity, LQG competes with string theory.[6]
Loop quantum gravity is an attempt to develop a quantum theory of gravity based directly on Einstein's geometric formulation rather than the treatment of gravity as a force. To do this, in LQG theory space and time are quantized analogously to the way quantities like energy and momentum are quantized in quantum mechanics. The theory gives a physical picture of spacetime where space and time are granular and discrete directly because of quantization just like photons in the quantum theory of electromagnetism and the discrete energy levels of atoms. An implication of a quantized space is that a minimum distance exists.
LQG postulates that the structure of space is composed of finite loops woven into an extremely fine fabric or network. These networks of loops are called spin networks. The evolution of a spin network, or spin foam, has a scale on the order of a Planck length, approximately $10^{-35}$ metres, and smaller scales are meaningless. Consequently, not just matter, but space itself, prefers an atomic structure.
The areas of research, which involves about 30 research groups worldwide, [7] share the basic physical assumptions and the mathematical description of quantum space. Research has evolved in two directions: the more traditional canonical loop quantum gravity, and the newer covariant loop quantum gravity, called spin foam theory.
The most well-developed theory that has been advanced as a direct result of loop quantum gravity is called loop quantum cosmology (LQC). LQC advances the study of the early universe, incorporating the concept of the Big Bang into the broader theory of the Big Bounce, which envisions the Big Bang as the beginning of a period of expansion that follows a period of contraction, which one could talk of as the Big Crunch.''\footnote{This paragraph is a direct quotation of ``Loop Quantum Gravity'' from
Wikipedis's contribution to that topic.}

 \subsection{The author's view of selected properties of \\Loop Quantum Gravity}
 The story that follows may not correctly follow the history, but the author offers a reasonable
 attempt, especially regarding the most important features.
 
 The basic elements for classical gravity are the metric $g_{ab}(x)=g_{ba}(x)>0$ and the 
 momentum $\pi^{cd}(x)=\pi^{dc}(x)$ \cite{adm}.
 The metric positivity complicates canonical quantization so it is modified by introducing 
 $E^i_a(x)\in I\!\!R^9$, with $a,b,c,..=i,j,k,..=1,2,3$, and set  $g_{ab}(x)=E^i_a(x)\,\delta_{ij}\,E^j_b(x)\geq0$. This is not strictly positive because $E^i_a(x)\in I\!\!R^9$, and $E^k_c(x) =0$ is not excluded as canonical quantization requires.
 It is evident that $E^i_a(x)$ is dimensionless to ensure that the metric is also dimensionless.
 
 A partner expression $A^a_i(x)$ is introduced to pair with $E^i_a(x)$, and these two variables lead to a Poisson bracket, effectively of unity, which is a key property for canonical quantization. The expression 
 $A^a_i(x)$ is not dimensionless, but, as  canonical quantization requires, $A^a_i(x)$ has the dimension of $\hbar$. Indeed, these expressions, when quantized, basically lead to
 $[\hat{E}^i_a(x), \hat{A}^b_j(x')] =i\hbar \,\delta^i_j\,\delta^b_a \,\delta^3(x,x') \,1\!\!1$.
 
 While we have begun the quantization study, it is essential to first ascertain whether   or not we have found physically correct quantum operators or not. This issue arises because while there are many different classical variables that are acceptable for some expression, e.g., the pair of variables
(skipping the usual indices)  $ F(E,A)=\ol{F}(\ol{E},\ol{A})$, but, when quantized,  $\hat{F}(\hat{E},\hat{A})\neq \hat{\ol{F}}(\hat{\ol{E}},\hat{\ol{A}})$. 
This fact implies, generally, there is only one set of quantum operators that is physically correct, and some rule must decide which are the classical functions that promote to physically correct quantum functions. For canonical quantization, Dirac \cite{dirac} suggested that physically correct quantum operators are those that were promoted from Cartesian coordinates. Dirac proposed this requirement, but did not prove his claim. Recently, the present author \cite{67} has proved that Dirac was correct, namely: using canonical quantization, `constant zero curvature' classical coordinates are those that lead to physically correct quantum operators.

The relevance of this paragraph is since the classical variables $E^i_a(x)$ and $A^b_j(x)$ are not Cartesian coordinates \cite{999}, the proposed canonical quantization does not have a physically  
correct quantization for the operators $\hat{E}_a^i(x)$ and $\hat{A}^b_j(x)$. 

The present canonical  gravity quantization leads to a very tiny,  but not zero, discrete behavior for space and time \cite{234}. The author attributes this proposed discrete behavior to a non-physical choice of basic quantum operators.  On the other hand, the present affine  gravity quantization, as we will show, has no discrete behavior, but instead, has a smooth and continuous behavior of space and time.

\subsubsection{Quantum gravity efforts}
The topic of quantum gravity has gathered numerous suggestions offered in Wikipedia \cite{234}, but no
mention of affine quantum gravity has appeared there.

 \section{An Affine Quantization of \\Einstein's Gravity}
\subsection{Favored variables for affine quantization}
   In an affine quantization our favored variables are the metric $g_{ab}(x)$, which is 
   dimensionless, 
   and the momentric
    $\pi^c_d(x) \;(\equiv \pi^{ce}(x)\,g_{de}(x))$, which has the
    dimension of $\hbar$.\footnote{Named for the  {\it momen}tum and the me{\it tric}.}
    The Poisson brackets become 
    \bn &&\hskip.1em \{\pi^a_b(x), \pi^c_d(y)\}=  (1/2)\,\s[\delta^a_d\s \pi^c_b(x)-\delta^c_b\s \pi^a_d(x)\s]\,\delta^3(x,y) \no \\
     &&\{g_{ab}(x), \pi^c_d(y)\} = (1/2)[\,\delta^c_a g_{bd}(x)+\delta^c_b g_{ad}(x)\,]\,
     \delta^3(x,y)  \no \\
     &&\hskip-.1em \{g_{ab}(x), g_{cd}(y)\}=0 \label{55} \;. \en
     
     In addition, we can promote these expressions to quantum operators, which then become
     \bn  &&[\hat{\pi}^a_b(x), \hat{\pi}^c_d(y)] = i(1/2)\hbar\,\s[\delta^a_d\s \hat{\pi}^c_b(x)-\delta^c_b\s 
     \hat{\pi}^c_a(x)]\,\delta^3(x,y) \no \\
     &&[\hat{g}_{ab}(x), \hat{\pi}^c_d(y)] =i(1/2)\,\hbar\,[\,\delta^c_a \hat{g}_{bd}(x)+\delta^c_b 
     \hat{g}_{ad}(x)\,]\delta^3(x,y) \no \\
    &&[\hat{g}_{ad}(x), \hat{g}_{cd}(y)] = 0 \;. \label{78} \en
    
    Observe that the middle line of (\ref{55}) -- and thus also that of  (\ref{78}) --   are still true if the metric variables change sign. This implies that using affine variables can lead to either strictly positive metrics, which we accept, or it could also lead to strictly negative metrics, which we reject. This important feature does not hold with analogous canonical classical and canonical quantum  equations.
    
    In the next section we will establish the fact that the classical Poisson brackets in (\ref{55}) match the  quantum commutation relations in (\ref{78}).

 \subsection{Affine coherent states for quantum gravity}
The basic affine quantum operators, $\hat{g}_{ab}(x)$ and $\hat{\pi}^c_d(x)$, are fundamental to 
the required coherent states, which are chosen \cite{cs} as
  \bn |\pi;g\>\equiv e ^{(i/\hbar)\tint \pi^{ab}(x)\hat{g}_{ab}(x)\,d^3\!x}
         e^{-(i/\hbar)\tint \eta^c_d(x)\hat{\pi}^d_c(x)\,d^3\!x} \,|\alpha\>\;, \en
         where $\{\eta(x)\}$ is a $3\times 3$ symmetric, arbitrary matrix, and $\{g(x)\}\equiv 
         e ^{\{\eta(x)\}}>0$. The fiducial vector $|\alpha\>$ 
         satisfies the equation $[(\hat{g}_{ab}(x)-\delta_{ab}\,1\!\!1) +i \hat{\pi}^c_d(x)/\alpha(x)\hbar]$ $|\alpha\>=0$, which establishes that $\<\alpha|\hat{g}_{ab}(x)|\alpha\>=\delta_{ab}$ and
         $\<\alpha|\hat{\pi}^c_ d(x)|\alpha\>=0$.
         
         For clarity, we offer a further description for a metric example given by \vskip.5em
         
          $\left( \begin{array}{ccc} e^{w_1(x)} & 0 & 0 \\ 0 & e^{w_2(x)} & 0 \\ 
	           0 & 0 & e^{w_3(x)} \end{array} \right) = \exp\left( \begin{array}{ccc} w_1(x)
	          & 0 & 0 \\ 0 & w_2(x) & 0 \\ 0 & 0 & w_3(x) \end{array} \right)$ \vskip.5em
	          
	 \hskip-1.5em that we denote as $W(x)=$ exp$(w(x))$. Next, we introduce $O(x)$ and its transpose  $O(x)^T$ which are arbitrary, $3\times 3$, orthogonal matrices such that $O(x)^T O(x)=O(x) \,O(x)^T\equiv 1\!\!1$. A typical $3\times 3$ orthogonal matrix, given as an example, is \vskip.5em

	  $\left( \begin{array} {ccc} 1 & 0 & 0 \\ 0& \cos(\th_1(x)) & -\sin(\th_1(x)) \\ 0 & \sin(\th_1(x)) & \cos(\th_1(x))  \end{array} \right) $  \vskip.5em

   \hskip-1.5em while changing $2$ more positions of ``1'' leads to $3$ angle parameters and to add $3$ more parameters from $W(x)$, leads to $6$ variables. 
   
   As expected, $6$ is the number of parameters in  $g_{ab}(x)=g_{ba}(x)>0$.

\subsection{The semi-classical gravity expression}

The Hamiltonian expressions for gravity are the most difficult part of the theory. According to ADM 
\cite{adm}, the classical canonical expression is given, using metric $g_{ab}(x)$ and momentum $\pi^{cd}(x)$ variables (with $\pi^{ce}(x) g_{de}(x)=\pi^c_d(x)$) as
   \bn H(\pi , g)=\tint \{ g^{-1/2}(x)[\pi^a_b(x)) \pi^b_a(x)-\half \pi^a_a(x)\pi^b_b(x)] + g^{1/2}(x)\,R(x)\} \;d^3\!x \;, \label{89} \en
   where $g(x)=\det[g_{ab}(x)]$, and $R(x)$ is the Ricci scalar term. Indeed, (\ref{89}) can also be considered to be written directly 
   in classical affine variables, namely $g_{ab}(x)$ and $\pi^c_d(x)$. In this case we rename
   the canonical expression, $H(\pi,g)$, to an affine expression, $H'(\pi,g)$.
   
   It is noteworthy that the momentum aspects of the classical Hamiltonian can be negative as well as positive. To simplify this aspect we adopt $2\times 2$ metrics. In that case, the first factors of
   (\ref{89}) could become $\pi^1_1=\pi^2_2=0, \pi^1_2=1$, and $\pi^2_1=-1$. In that case, 
   $\pi^a_b\,\pi^b_a =-2$. For this example, the second factors of (\ref{89})  $\pi^a_a\,\pi^b_b=0$. In addition, the Ricci scalar can be negative as well as positive. These properties carry over to the quantized elements, which could imply, a simple toy model, roughly like  $\H =P_1P_2+ Q_1Q_2$.

   The affine quantized version of expression (\ref{89}) using 
   metric, $\hat{g}_{ab}(x)$, and momnentric operators, $\hat{\pi}^c_d(x)$, is given by 
	   \bn &&\hskip-2em \H'(\hat{\pi}, \hat{g}) \label{777} \\ &&= \tint \{ [\hat{\pi}^a_b(x) \hat{g}^{-1/2}(x)\hat{\pi}^b_a(x)-\half \hat{\pi}^a_a(x)\hat{g}^{-1/2}(x)\hat{\pi}^b_b(x)] + 
	   \hat{g}^{1/2}(x)\,\hat{R}(x) \}\;d^3\!x \;. \no   \en
	   
	   Moreover, the connection between the classical and quantum terms arises from
	   \bn &&\hskip-3em H'(\pi,g)=\<\pi;g| \H'(\hat{\pi}, \hat{g}) |\pi;g\> \\ \no
          &&\hskip.9em =\<\alpha| \H'(\hat{\pi}^a_b+\pi^{ac}[e^{\eta/2}]_c^e \gg_{ed}[e^{\eta/2}]^d_b,
          [e^{\eta/2}]_c^a\gg_{ab}[e^{\eta/2}]^b_d) ]|\alpha
          \>\;, \no \\ 
          &&\hskip.9em  = \H'(\<\alpha|\hat{\pi}^a_b|\alpha\>+\pi^{ac}[e^{\eta/2}]_c^e
          \<\alpha|\gg_{ed}|\alpha\>[e^{\eta/2}]^d_b, [e^{\eta/2}]_c^a\<\alpha|\gg_{ab}|\alpha\>[e^{\eta/2}]^b_d )\; \no \\
          &&\hskip14em +\,\Op(\hbar;\pi,g) \;, \no \en
          which becomes
          \bn H'(\pi^a_b,g_{cd})=\H'(\pi^a_b,g_{cd})+\Op(\hbar; \pi,g) \en
          leading to the 
   usual classical limit as $\hbar\ra0$, or when $\Op$ is tiny enough to ignore. This expression confirms the equality of (\ref{55}) and (\ref{78}), as to the similarity of classical and quantum expressions.

\section{Schr\"odinger's Representation and Equation}
 In the previous section we developed the quantum Hamiltonian in (\ref{777}), and now we develop
 a specific formulation wherein the dimensionless metric operator is given by  $\gg_{ab}(x)\ra g_{ab}(x)$  and the dimensionless metric determinant is given by $\gg(x)\equiv \det[\gg_{ab}(x)]\ra g(x) \equiv \det[g_{ab}(x)]$, as well as 
 the momentric, which has the dimensions of $\hbar$, and offers a formal expression given by
  \bn \hat{\pi}^c_d(x)\ra -i(1/2)\hbar[g_{de}(x)(\d/\d g_{ce}(x))+(\d/\d g_{ce}(x))g_{de}(x)]\;.\en
  It is noteworthy that
         \bn \hat{\pi}^a_b(x) \,g(x)^{-1/2}=0 \label{888} \;, \en
   which is obtained by
          $ \hp^a_b \s F(g)=0$, and implies that
   $[g_{bc}\s (\d/\d g_{ac})+\half\delta^a_b]\s F(g)=0$. Next, we find  $g_{bc}\s g^{ac}\s g\s\s dF(g)/d g + \half\delta^a_b\s F(g)=0$, which requires that
   $ g\s d\s F(g)/d g+\half\s F(g)=0$; hence $F(g)\propto g^{-1/2}$.
   
        
  These expressions permit us to offer Schr\"odinger's equation, which is
   \bn i\hbar (\d\;\Psi(x, t)/\d t)&&\hskip3em  \no \\ &&\hskip-9em
   = \{\!\{ \tint [\![\hat{\pi}^a_b(x,t) g(x,t)^{-1/2}\hat{\pi}^b_a(x,t)
   -(1/2)\hat{\pi}^a_a(x,t) g(x,t)^{-1/2}\hat{\pi}^b_b(x,t) \no \\ 
   &&\hskip4em   +g(x,t)^{1/2}\;R(x,t)]\!]\;d^3\!x\}\!\}\;\Psi(x,t) \;.\en
   
   \subsection{Possible solutions to Schr\"oedinger equation}
   The relation in (\ref{888}) also leads to $\hat{\pi}^a_b(x)\;\Pi_y g(y)^{-1/2}=0$ for any 
   and all $y$. This property offers eigenfunctions having the general 
   form $\Psi(x)=V(x) \,\Pi_y g(y)^{-1/2}$; this expression is suggested here because in selected ultralocal procedures \cite{ul}, such behavior is required. The positivity and negativity of the Hamiltonian 
   operator points to continuum solutions rather than discrete levels of eigenfunctions. The 
   Hamiltonian is really a constraint that requires solutions of the form $\H'(\hat{\pi}(x),\gg(x))\,\Psi(x)
   =0$. In particular, the constraint expressions can also be correctly extracted using a suitable functional integration \cite{proj}.  This procedure points toward the physical features that are 
   the most difficult with the Hamiltonian operator, and simpler issues can complete the story
 which have been examined before in \cite{now}.

\end{document}